\documentclass[reprint,aps,prl,amsmath,amssymb]{revtex4-1}

\usepackage[utf8]{inputenc}
\usepackage{graphicx}
\usepackage{mathtools,bm}

\usepackage{braket}
\newcommand*{\expect}[1]{\braket{#1}}

\newcommand*{\commutator}[2]{\mathinner{[{#1},{#2}]}}

\newcommand*{\PP}[1]{\frac{\partial}{\partial{#1}}}

\newcommand*{\mat}[1]{\bm{\mathrm{#1}}}

\begin{document}

\title{Time-local Heisenberg--Langevin equations and the driven qubit}

\author{S. J. Whalen}
\email{simon.whalen@gmail.com}

\author{H. J. Carmichael}

\affiliation{The Dodd-Walls Centre for Photonic and Quantum Technologies, Department of Physics, University of Auckland, Private Bag 92019, Auckland, New Zealand}

\begin{abstract}
The time-local master equation for a driven boson system interacting with a boson environment is derived by way of a time-local Heisenberg--Langevin equation. Extension to the driven qubit fails---except for weak excitation---due to the lost linearity of the system-environment interaction. We show that a reported  time-local master equation for the driven qubit is incorrect. As a corollary to our demonstration, we also uncover odd asymptotic behavior in the ``repackaged'' time-local dynamics of a system driven to a far-from-equilibrium steady state: the density operator becomes steady while time-dependent coefficients oscillate (with periodic singularities) forever.
\end{abstract}

\maketitle

Treatments in the Heisenberg picture of a system in interaction with an environment date to the work, from the  1960s, by Senitzky \cite{senitzky_1960&61}, who considered the damped harmonic oscillator with a mind to applications to the radiation field in a cavity. The associated equations of motion subsequently took the name of quantum Langevin or Heisenberg--Langevin equations, recognizing the role of environment operators as a random force, or noise, as in the classical Langevin equation. Although this random force is never Markovian \cite{ford_etal_1988}, Markovian models can serve as an excellent approximation \cite{lax_2000}, and are widely used, particularly in quantum optics, where the input-output theory of Gardiner and Collett \cite{gardiner&collett_1985} is now canonical.

Issues of formal exactness aside, the current move from given quantum systems---an atom or radiation field in a cavity---to engineered systems, drives a more pragmatic interest in the theory of non-Markovian open quantum systems. While generalizations of input-output theory to the non-Markovian regime have been considered \cite{ciuti&carusotto_2006,diosi_2012,zhang_etal_2013}, more commonly the Schr\"odinger picture is adopted, where, after the work of Hu \emph{et al}. \cite{hu_etal_1992}, time-local master equations are derived
\cite{breuer_etal_1999,anastopoulos&hu_2000,cresser_2000,strunz&yu_2004,an&zhang_2007,xiong_etal_2010,lei&zhang_2012,tu&zhang_2008}. Of particular interest in this paper are the time-local master equation for a driven boson system in interaction with a boson environment (first derived in \cite{cresser_2000}) and the equation for spontaneous emission from a two-state system, or qubit \cite{breuer_etal_1999,anastopoulos&hu_2000,cresser_2000,strunz&yu_2004}. They invite a conflation: a time-local master equation for the driven qubit.

The driven qubit is an important example, considering its role in quantum information science and the numerous physical realizations. The question of a time-local master equation is an old one. It is raised in Sec.\ IVA of Ref.~\cite{cresser_2000}, where, after first noting obstacles to its derivation, the author mentions an equation reported in a preprint \cite{anastopoulos&hu_1999}, following with: ``Such a relatively simple result seems to be inconsistent with the conclusion reached above about the difficulty of dealing with the two level atom problem, and is hence an issue that requires further consideration.'' The noted equation is absent from the published version of \cite{anastopoulos&hu_1999} (Ref.~\cite{anastopoulos&hu_2000}), which can be seen to endorse the call for ``further consideration.''

The ``relatively simple result'' substitutes qubit raising and lowering operators for the creation and annihilation operators in the time-local master equation for the driven boson system. Having appeared first in \cite{anastopoulos&hu_1999}, it reappears in a recent work of Shen \emph{et al}.\ \cite{shen_etal_2014}, along with a derivation from Feynman-Vernon influence functional theory which invokes a coherent state representation in Grassmannian variables for the qubit state. We show in this paper that the derived time-local master equation is incorrect; it is not influence functional theory and coherent-state path integrals that yield a successful derivation, but linearity, which for the driven qubit is lost. While in spontaneous emission \cite{breuer_etal_1999,anastopoulos&hu_2000,cresser_2000,strunz&yu_2004} linearity is effectively retained---due to the one-quantum truncation---for the driven qubit, multi-photon scattering (nonlinearity) becomes important.

As introduction and background, we first recover the known time-local master equation for the driven boson system \cite{cresser_2000,lei&zhang_2012} from a time-local Heisenberg--Langevin equation. Our use of the Heisenberg picture exposes the essentials of a successful derivation and underpins the discussion of the driven qubit that follows.

A system $S$ and environment $E$ have free Hamiltonian $H_0=\sum_j\omega_ja_j^\dagger a_j+\sum_{\alpha}\sum_{l_\alpha}\omega_{l_\alpha}b_{\alpha l_\alpha}^\dagger b_{\alpha l_\alpha}$, where $a_j$ ($b_{\alpha l_\alpha}$) and $\omega_j$ ($\omega_{l_\alpha}$) are system (environment) operators and frequencies; the index $j$ ($\alpha$) labels different subsystems (subenvironments) and the index $l_\alpha$ labels the modes of subenvironment $\alpha$.
Operators of $S$ and $E$ commute,
while  $\commutator{a_j}{a_{j^\prime}^\dagger}=\delta_{jj^\prime}$, $\commutator{b_{\alpha l}}{b_{\alpha^\prime l^\prime}^\dagger} = \delta_{\alpha\alpha^\prime} \delta_{ll^\prime}$. We adopt the rotating-wave approximation, such that in an interaction picture generated by $H_0-\sum_j\Delta_{jj}a_j^\dagger a_j$, $\Delta_{jj}=\omega_j-\omega_0$, the Hamiltonian of $S$ in interaction with $E$ is
\begin{equation}
 H(t)=H_{S}(t)+H_{SE}(t),
 \label{eqn:hamiltonian}
\end{equation}
where
\begin{equation}
 H_{S}(t)=\sum_{jk}\Delta_{jk}a_j^\dagger a_k+\sum_j\left[ \Omega_j(t)a_j^\dagger+\Omega_j^*(t)a_j\right],
 \label{eqn:system_hamiltonian}
\end{equation}
with $\Delta_{jk}^*=\Delta_{kj}$, accounts for detuning ($\Delta_{jj}$), coupling among subsystems ($\Delta_{jk}, k\neq j$), and driving ($\Omega_j$), and
\begin{equation}
 H_{SE}(t)=\sum_j\big[a_j^\dagger B_j(t)+B_j^\dagger(t)a_j\big],
 \label{eqn:interaction_hamiltonian}
\end{equation}
with
\begin{equation}
B_j(t)=\sum_{\alpha l_\alpha}\kappa_{j\alpha l_\alpha}\exp[i(\omega_0-\omega_{l_\alpha})t]b_{\alpha l_\alpha},
\label{eqn:noise_operator}
\end{equation}
where $\kappa_{j \alpha l_\alpha}$ are coupling constants. We assume the state of the entire system to be separable, $\rho = \rho_S \otimes \rho_E$, with subenvironment $\alpha$ in a thermal state of temperature $T_\alpha$, i.e.,
$\rho_E=Z^{-1}\prod_{\alpha}\exp\big(-T_{\alpha}^{-1}\sum_{l_\alpha}\omega_{l_\alpha}b_{\alpha l_\alpha}^\dagger b_{\alpha l_\alpha}\big)$, where $Z$ denotes the partition function. We also introduce the dissipation kernel
\begin{align}
F_{jk}(t_1-t_2)&=\commutator{B_j(t_1)}{B_k^\dagger(t_2)}\notag\\
\noalign{\vskip6pt}
&=\sum_{\alpha} \int_{-\infty}^\infty d{\omega}J_{\alpha jk}(\omega)e^{-i\omega (t_1-t_2)},
\label{eqn:dissipation_kernel}
\end{align}
and the noise kernel
\begin{align}
G_{jk}(t_1-t_2)&=\expect{B_k^\dagger(t_2) B_j(t_1)}\notag\\
\noalign{\vskip6pt}
&=\sum_{\alpha} \int_{-\infty}^\infty d{\omega}J_{\alpha jk}(\omega)n_\alpha(\omega)e^{-i\omega(t_1-t_2)},
 \label{eqn:noise_kernel}
\end{align}
with
\begin{equation}
J_{\alpha jk}(\omega)=\sum_{l_\alpha}\kappa_{j \alpha l_\alpha}\kappa_{k \alpha l_\alpha}^*\delta(\omega+\omega_0-\omega_{l_\alpha})
\end{equation}
a spectral density and $n_\alpha(\omega)=\{\exp[(\omega+\omega_0)/T_\alpha]-1\}^{-1}$ the Bose--Einstein distribution for subenvironment $\alpha$.

Formally eliminating the environment operators $B_j(t)$ from the Heisenberg equation of motion for system operator $A(t)$ yields the Heisenberg--Langevin equation:
\begin{widetext}
\begin{align}
\frac d{dt}A(t)=&\,\,\PP{t}A(t)-i\commutator{A(t)}{H_{S}(t)}+\sum_{jk} \int_0^tdt^\prime\left\{F_{jk}(t-t^\prime)\commutator{a_j^\dagger(t)}{A(t)}a_k(t^\prime)\right.\notag\\
\noalign{\vskip4pt}
&\left.+F_{jk}^*(t-t^\prime)a_k^\dagger(t^\prime)\commutator{A(t)}{a_j(t)}\right\}+i\sum_j
\left\{\commutator{a_j^\dagger(t)}{A(t)}B_j(t)-B_j^\dagger(t)\commutator{A(t)}{a_j(t)}\right\},
\label{eqn:heisenberg-langevin_equation}
\end{align}
\end{widetext}
with alternate compact form
\begin{align}
\frac d{dt}A((t)=&\,\,\PP{t}A(t)-\sum_j\left\{[a_j^\dagger(t),A(t)]\left(\frac d{dt}a_j(t)\right)\right.\notag\\
&\left.+\left(\frac d{dt}a_j^\dagger(t)\right)[A(t),a_j(t)]\right\}.
\label{eqn:heisenberg_equations_connected}
\end{align}
Thus if the derivatives $da_j(t)/dt$ can be written in time-local form, so can the Heisenberg--Langevin equation.

To this end, we collect the $a_j(t)$ into the column vector, ${\bm a}(t)$, and similarly introduce ${\bm\Omega}(t)$, ${\bm B}(t)$, and matrices ${\bm\Delta}$ and ${\bm F}(t_{12})$. With this notation, from Eq.~(\ref{eqn:heisenberg-langevin_equation}),
\begin{align}
\frac d{dt}{\bm a}(t)=&-i{\bm\Delta}{\bm a}(t)-\int_0^tdt^\prime{\bm F}(t - t^\prime){\bm a}(t^\prime)\notag\\
\noalign{\vskip4pt}
&-i[{\bm\Omega}(t)+{\bm B}(t)],
\label{eqn:vector&time-nonlocal_form}
\end{align}
which we solve by Laplace transforms:
\begin{equation}
{\bm a}(t) = {\bm V}(t){\bm a}-i({\bm V}*{\bm\Omega})(t)-i{\bm C}(t),
\label{eqn:solution}
\end{equation}
with ${\bm C}(t)=({\bm V}\ast{\bm B})(t)$, where $\ast$ denotes convolution and the Green's function, ${\bm V}(t)$, satisfies
\begin{equation}
\frac d{dt}{\bm V}(t)=-i{\bm\Delta}{\bm V}(t)-\int_0^tdt^\prime{\bm F}(t-t^\prime){\bm V}(t^\prime),
\label{eqn:green_function}
\end{equation}
with ${\bm V}(0)=\mat{I}_{N}$, $N$ the range of index $j$.

Equations~(\ref{eqn:solution}) and (\ref{eqn:green_function}) constitute a solution which might be used to evaluate expectations and correlation functions directly. Our aim, though, is to present it as an equation of motion in time-local form. We therefore differentiate Eq.~\eqref{eqn:solution} and then use the same equation to eliminate the initial condition. Assuming ${\bm V}(t)$ invertible, and introducing
\begin{equation}
{\bm\gamma}(t)=-[d{\bm V}(t)/dt]{\bm V}(t)^{-1},
\label{eqn:time-dependent_gamma}
\end{equation}
we thereby arrive at
\begin{equation}
\frac d{dt}{\bm a}(t)=-{\bm\gamma}(t){\bm a}(t)-i{\bm \xi}(t)-i{\bm D}(t),
\label{eqn:vector&time-local_form}
\end{equation}
with
\begin{equation}
{\bm\xi}(t)=[{\bm\gamma}(t)+d/dt]({\bm V}\ast{\bm\Omega})(t)
\end{equation}
and ${\bm D}(t)=[{\bm\gamma}(t)+d/dt]{\bm C}(t)$. Equations~(\ref{eqn:heisenberg_equations_connected}) and (\ref{eqn:vector&time-local_form}) cast the Heisenberg--Langevin equation, Eq.~(\ref{eqn:heisenberg-langevin_equation}), into time-local form:
\begin{widetext}
\begin{align}
\frac d{dt}A(t)=&\,\,\PP{t}A(t)-i\sum_j\commutator{A(t)}{\xi_j(t)a_j^\dagger(t)+\xi_j^*(t)a_j(t)}+\sum_{jk}\left\{\gamma_{jk}(t) \commutator{a_j^\dagger(t)}{A(t)}a_k(t)\right.\notag\\
&\left.+\gamma_{jk}^*(t)a_k^\dagger(t)\commutator{A(t)}{a_j(t)}\right\}+i\sum_j\left\{\commutator{a_j^\dagger(t)}{A(t)}D_j(t)-D_j^\dagger(t) \commutator{A(t)}{a_j(t)}\right\}.
\label{eqn:heisenberg-langevin_equation_time-local_form}
\end{align}
\end{widetext}

We move to a master equation by taking expectation values on both sides of Eq.~\eqref{eqn:heisenberg-langevin_equation_time-local_form}, where only the last term on the right involves operators of the environment. Wick's theorem allows us to deal with the one troublesome term: for Gaussian $\rho_E$, as below Eq.~(\ref{eqn:noise_operator}), it allows us to factor out contractions of $C_k^\dagger(t)$---entering from the adjoint of Eq.~(\ref{eqn:solution})---and $D_j(t)$; specifically, we find
\begin{equation}
\expect{A(t)D_j(t)}=i\sum_k \expect{C_k^\dagger(t)D_j(t)}\expect{\commutator{a_k(t)}{A(t)}},
\label{eqn:wick's_theorem}
\end{equation}
for any $A(t)$ that satisfies Eq.~\eqref{eqn:heisenberg-langevin_equation_time-local_form}. Using this result, the expectation of the last term on the right in Eq.~\eqref{eqn:heisenberg-langevin_equation_time-local_form} may be expanded as the sum over expectations of system operators alone, $\sum_{jk}\lambda_{jk}(t)\langle[[a_j^\dagger(t),A(t)],a_k(t)]\rangle$,  where
\begin{equation}
\lambda_{jk}(t)=\expect{C_k^\dagger(t) D_j(t)}+\expect{D_k^\dagger(t) C_j(t)}
\end{equation}
is evaluated from
\begin{equation}
{\bm \lambda}(t)=\frac d{dt}{\bm W}(t)+{\bm\gamma}(t){\bm W}(t)+{\bm W}(t){\bm\gamma}^\dagger(t),
\end{equation}
with the matrix ${\bm W}(t)$, $W_{jk}(t)=\expect{C_k^\dagger(t)C_j(t)}$ [with ${\bm C}(t)$ defined below Eq.~(\ref{eqn:heisenberg-langevin_equation})], evaluated from Eq.~(\ref{eqn:noise_kernel}):
\begin{equation}
{\bm W}(t)=\int_0^tdt_1\int_0^tdt_2{\bm V}(t-t_1){\bm G}(t_{12}){\bm V}^\dagger(t-t_2).
\end{equation}
With all expectations now system operator expectations, the time-local master equation can be read from the time-local Heisenberg--Langevin equation by inspection:
\begin{widetext}
\begin{align}
\frac d{dt}\rho_S(t)=&-i\sum_j\commutator{\xi_j(t) a_j^\dagger+\xi_j^*(t)a_j}{\rho_S(t)}+\sum_{jk}\left\{\gamma_{jk}(t)\commutator{a_k \rho_S(t)}{a_j^\dagger}+\gamma_{jk}^*(t)\commutator{a_j}{\rho_S(t) a_k^\dagger} \right\}\notag\\
&+\sum_{jk}\lambda_{jk}(t)\commutator{\commutator{a_k}{\rho_S(t)}}{a_j^\dagger}.
\label{eqn:time-local_master_equation}
\end{align}
\end{widetext}
The result agrees with that obtained from the widely used Feynman-Vernon influence functional method by Lei and Zhang \cite{lei&zhang_2012}; and for $T_\alpha=0$ with Cresser \cite{cresser_2000}.

The treatment above may be extended by adding an interaction term $\sum_{jk}(\chi_{jk}a_j^\dagger a_k^\dagger+{\rm h.c.})$, with $\chi_{jk}$ a complex constant, to $H_S(t)$. The evolution is still linear, but with annihilation and creation operators coupled. Parallel to Eq.~(\ref{eqn:vector&time-nonlocal_form}), one  sets out the equations of motion as a vector equation of twice the dimension whence the time-local form, as in Eq.~(\ref{eqn:vector&time-local_form}), follows.
It is also possible to include counter-rotating terms in the interaction between system and environment, in which case it is natural to omit the
transformation to the interaction picture. Chang and and Law \cite{chang&law_2010} have solved the Heisenberg--Langevin equation for one subsystem with a parametric interaction keeping counter-rotating terms. They reach a time-local master equation by assuming a Gaussian state and matching the equations of motion for moments up to second order. The method is similar to ours, though a Heisenberg--Langevin equation in time-local form,  Eq.~(\ref{eqn:heisenberg-langevin_equation_time-local_form}), is not considered. Note that Eq.~(\ref{eqn:heisenberg-langevin_equation_time-local_form}) is specific to the Hamiltonian adopted.

We now turn to the driven qubit and begin by asking whether the above generalizes to this apparently related case. We limit ourselves to one subsystem for simplicity's sake, i.e., we consider the Hamiltonian
\begin{equation}
H_{S}(t)=\Delta \sigma_+\sigma_-+\Omega(t)\sigma_++\Omega^*(t)\sigma_-,
\label{eqn:system_hamiltonian_qubit}
\end{equation}
where $\sigma_+$ and $\sigma_-$ are raising and lowering operators, with commutation relation $[\sigma_+,\sigma_-]=\sigma_z$. In place of Eq.~(\ref{eqn:vector&time-nonlocal_form}), we have the Heisenberg--Langevin equation for the qubit lowering operator
\begin{align}
\frac d{dt}\sigma_-(t)=&\,\,\sigma_z(t)\left\{i\Delta \sigma_-(t)+\int_0^tdt^\prime F(t-t^\prime)\sigma_-(t^\prime)\right.\notag\\
&\left.+i[\Omega(t)+B(t)]\vphantom{\int}\right\},
\label{eqn:heisenberg-langevin_equation_qubit}
\end{align}
and in place of Eq.~(\ref{eqn:heisenberg_equations_connected}),
\begin{align}
\frac d{dt}A(t)=&\,\,\PP{t}A(t)+\commutator{\sigma_+(t)}{A(t)}\sigma_z(t)\left(\frac d{dt}\sigma_-(t)\right)\notag\\
&+\left(\frac d{dt}\sigma_+(t)\right)\sigma_z(t) \commutator{A(t)}{\sigma_-(t)}.
\label{eqn:heisenberg_equations_connected_qubit}
\end{align}
Thus, if we are able to express the derivative of $\sigma_-(t)$ in time-local form, we can use Eq.~(\ref{eqn:heisenberg_equations_connected_qubit}) to derive a time-local Heisenberg--Langevin equation and corresponding master equation. Unfortunately, Eq.~(\ref{eqn:heisenberg-langevin_equation_qubit}) is nonlinear and not so easily solved. In fact, to our knowledge, no closed-form solution exists---even in the Markovian case. We cannot, then, obtain time-local Heisenberg--Langevin and master equations for the driven qubit.

Suppose, though, that the qubit is only very weakly excited, so that at all times it remains close to its ground state. In this case we may approximate $\sigma_z(t)$ by $-1$, and Eq.~(\ref{eqn:heisenberg-langevin_equation_qubit}) reads as Eq.~(\ref{eqn:vector&time-nonlocal_form}) with $a(t)\to\sigma_-(t)$; the above procedure then leads to Eqs.~(\ref{eqn:heisenberg-langevin_equation_time-local_form}) and (\ref{eqn:time-local_master_equation}) with qubit rasing (lowering) operators standing in for boson creation (annihilation) operators. We emphasize, however, that the simple extrapolation holds only when the driving is weak. Shen \emph{et al}.~\cite{shen_etal_2014} report a derivation with no such restriction; they assume a zero-temperature environment but admit coherent driving of any strength. They return to the equation from \cite{anastopoulos&hu_1999} and revive Cresser's concern  \cite{cresser_2000}: ``Such a relatively simple result $\mkern-2mu\ldots\mkern-2mu$ is an issue that requires further consideration.''

We answer the concern below by demonstrating that the extrapolation from the driven boson system to the driven qubit is, most generally, incorrect. It is known to hold, however, if there is no drive at all \cite{breuer_etal_1999,cresser_2000,anastopoulos&hu_2000,strunz&yu_2004}. Since the latter follows despite the nonlinearity in Eq.~(\ref{eqn:heisenberg-langevin_equation_qubit}), we first set $\Omega(t)$ to zero and show why this is so.

Consider the initial state to be an arbitrary state of the qubit with the environment in the vacuum state, i.e., $|\psi(0)\rangle=\alpha|g,0\rangle+\beta(0)|e,0\rangle$, $|\alpha|^2+|\beta(0)|^2=1$, where $g$ ($e$) denotes the ground (excited) state of the qubit, and $0$ denotes the environment vacuum state. Note then that the global ground state, $|g,0\rangle$, does not evolve, while the qubit emits at most one photon; thus, the state at a later time admits the truncated expansion
\begin{equation}
|\psi(t)\rangle=\alpha|g,0\rangle+\beta(t)|e,0\rangle+\sum_l\eta_l(t)|g,l\rangle,
\label{eqn:one-quantum_expansion}
\end{equation}
where $l$ labels the mode of the environment occupied by the single photon. From this expansion alone, the trivial identity $d\beta(t)/dt=-\gamma(t)\beta(t)$, $\gamma(t)=-[d\beta(t)/dt]\beta(t)^{-1}$, yields the time-local master equation
\begin{equation}
\frac d{dt}\rho_S(t)=\gamma(t)[\sigma_-\rho_S(t),\sigma_+]+\gamma^*(t)[\sigma_-,\rho_S(t)\sigma_+].
\label{eqn:time-local_master_equation_spontaneous_emission}
\end{equation}
It remains to find an explicit expression for $\gamma(t)$. To this end, we place Eq.~(\ref{eqn:heisenberg-langevin_equation_qubit}) between $\langle g,0|$ and $|\psi(0)\rangle$, and use $B(t)|\psi(0)\rangle=0$ and
\begin{align}
\langle g,0|\sigma_z(t)\sigma_-(t^\prime)|\psi(0)\rangle&=-\langle g,0|\sigma_-|\psi(t^\prime)\rangle\notag\\
\noalign{\vskip2pt}
&=-\beta(t^\prime),
\label{eqn:Heisenberg_to_Schroedinger}
\end{align}
where the latter result follows by transferring the unitary evolution from operators to states---again, $|g,0\rangle$ does not evolve, and $\sigma_z|g,0\rangle=-|g,0\rangle$. We find that $\beta(t)$ satisfies Eq.~(\ref{eqn:green_function}), whence $\gamma(t)=-[dV(t)/dt]V(t)^{-1}$.

The requirement that $|g,0\rangle$ not evolve is a precondition of the one-photon truncation and central to this derivation. It is only met for zero drive; otherwise the equation derived in Ref.~\cite{shen_etal_2014}, where $-i\commutator{\xi(t)\sigma_++\xi^*(t)\sigma_-}{\rho_S(t)}$ is added to the right-hand side of Eq.~(\ref{eqn:time-local_master_equation_spontaneous_emission}), is incorrect. We can demonstrate its failure by considering the Lorentzian spectral density,
\begin{equation}
J(\omega)=\frac{\Gamma}{2\pi}\frac{\lambda^2}{{(\omega-\Delta+\delta)}^2+\lambda^2},
\label{eqn:lorentzian_spectral_density}
\end{equation}
from Sec.~VB of that paper, with $\Delta$ ($\delta$) the detuning of the qubit from the drive (center of the Lorentzian line), $\lambda$ the Lorentzian linewidth, and $\Gamma$ a parameter to control the system-environment interaction strength; we adopt a frame rotating with the drive and take $\Omega(t)$ constant.

The dissipation kernel and Green's function [Eqs.~(\ref{eqn:dissipation_kernel}) and (\ref{eqn:green_function})] can be found analytically for the Lorentzian spectral density. The master equation derived in Ref.~\cite{shen_etal_2014} is then fully defined by $\gamma(t)$, above, and $\xi(t)=[\gamma(t)+d/dt](V*\Omega)(t)$, with $V(t)$ the solution to Eq.~(\ref{eqn:green_function}); with it we can directly compute the driven qubit response. Alternatively, the Lorentzian spectral density maps to a damped harmonic oscillator initially in the vacuum state. Thus, we can check the response against $\rho_S(t)={\rm tr}_b[\rho(t)]$, with the trace taken over the harmonic oscillator state and $\rho(t)$ obeying the master equation
\begin{equation}
\frac d{dt}\rho(t)=-i\commutator{H}{\rho(t)}+{\mathcal L}_b[\rho(t)],
\label{eqn:master_equation_lorentzian_spectrum}
\end{equation}
where ${\mathcal L}_b[\,\cdot\,]=\lambda\left[2b\cdot b^\dagger-b^\dagger b\cdot-\cdot b^\dagger b\right]$ and
\begin{align}
H=&\,\,\Delta\sigma_+\sigma_-+(\Delta-\delta)b^\dagger b+\Omega\left(\sigma_++\sigma_-\right)\notag\\
&+(\lambda\Gamma/2)^{1/2}\left(\sigma_+b+b^\dagger\sigma_-\right).
\label{eqn:system_hamiltonian_lorentzian_spectrum}
\end{align}

\begin{figure}[t]
\begin{center}
\includegraphics[width=3.4in,height=3.0in]{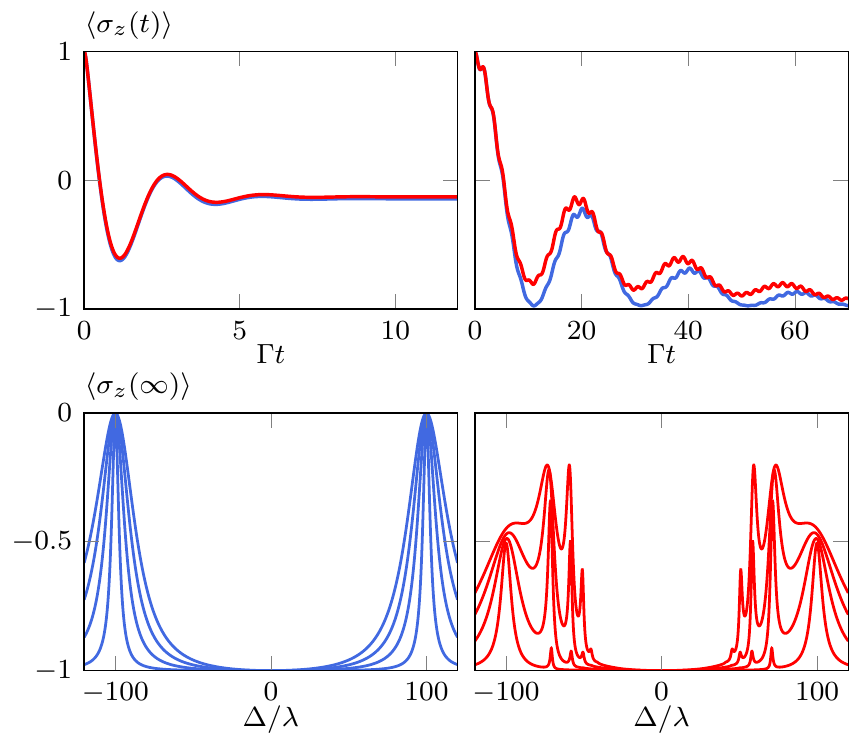}
\end{center}
\vskip-0.5cm
\caption{Comparison of the time-local master equation of Ref.~\cite{shen_etal_2014} with the equivalent model, Eq.~(\ref{eqn:master_equation_lorentzian_spectrum}), for a Lorentzian spectral density. (Top) Evolution of $\expect{\sigma_z(t)}$ to steady state, for $\lambda = 25$, $\Delta = 0.3$, $\Omega = 1$, $\delta = 0.01$ (Left), and $\lambda = 0.05$, $\Delta = 3.5$, $\Omega = 0.4$, $\delta = 0.01$ (Right)---parameters in units of $\Gamma$; from Ref.~\cite{shen_etal_2014} (blue) and Eq.~(\ref{eqn:master_equation_lorentzian_spectrum}) (red). (Bottom) Steady-state response $\expect{\sigma_z(\infty)}$ versus $\Delta$ from Eq.~(\ref{eqn:steady-state_response2}) (Left) and Eq.~(\ref{eqn:master_equation_lorentzian_spectrum}) (Right); for $(\Gamma/2\lambda)^{1/2}=100$ and $\Omega/\lambda=4$, $10$, $16$, $22$ (lower curve to upper curve).}
\label{fig:fig1}
\end{figure}

The upper two frames of Fig.~\ref{fig:fig1} reproduce the curves from Figs.~2(b) and 5(b) of Ref.~\cite{shen_etal_2014}, comparing results computed from Eq.~(\ref{eqn:master_equation_lorentzian_spectrum}). The frame on the left displays decay from the excited state in the presence of relatively strong driving, where, since $\lambda=25\Gamma$ brings us close to the Markov limit, the agreement, as expected, is good. On the right the driving is weaker and the drive detuning greater; nonetheless, although the excitation in steady-state is less, the decay takes place away from the Markov limit and a clear discrepancy has set in. More generally, the disagreement can be far more substantial, even of a qualitative nature, as we now show by considering solutions in the asymptotic limit $t\to\infty$.

The time-local master equation is equivalent to Bloch equations with time-dependent coefficients, $\gamma(t)$ (decay and detuning) and $\xi(t)$ (drive). These equations admit a steady-state solution, $\expect{\sigma_-(\infty)}$ and $\expect{\sigma_z(\infty)}$ independent of time, if
\begin{align}
\gamma(t)\expect{\sigma_-(\infty)}&=i\xi(t)\expect{\sigma_z(\infty)},
\label{eqn:bloch1}\\
\noalign{\vskip2pt}
{\rm Re}\{\gamma(t)\}[\expect{\sigma_z(\infty)}+1]&={\rm Re}\{2i\xi^*(t)\expect{\sigma_-(\infty)}\}.
\label{eqn:bloch2}
\end{align}
There is no guarantee that $\gamma(t)$ and $\xi(t)$ reach a steady value, but the ratio $\gamma(t)^{-1}\xi(t)=[1+\gamma(t)^{-1}d/dt](V\ast\Omega)(t)$ must because $V(t)$ decays to zero---its convolution with constant $\Omega$ is an integral of fixed asymptotic value. Equations~(\ref{eqn:bloch1}) and (\ref{eqn:bloch2}) then give
\begin{equation}
\expect{\sigma_z(\infty)}=-\left[1+2\left|\lim_{t\to\infty}(V\ast\Omega)(t)\right|^2\right]^{-1},
\label{eqn:steady-state_response1}
\end{equation}
and from the asymptotic value of the convolution, setting $\delta=0$ for simplicity,
\begin{equation}
\expect{\sigma_z(\infty)}=-\left[1+2\frac{\Omega^2(\lambda^2+\Delta^2)}{(\lambda\Gamma/2-\Delta^2)^2+\lambda^2\Delta^2}\right]^{-1}.
\label{eqn:steady-state_response2}
\end{equation}

Consider the coupling $(\lambda\Gamma/2)^{1/2}$ much larger than the linewidth $\lambda$; Eq.~(\ref{eqn:steady-state_response2}) clearly describes a response peaked at $\Delta=\pm(\lambda\Gamma/2)^{1/2}$. This doublet follows from vacuum Rabi splitting and is just what we would expect. What is not expected, though, is the featureless power broadening shown in the bottom-left panel of Fig.~\ref{fig:fig1}. The response computed from  Eq.~(\ref{eqn:master_equation_lorentzian_spectrum}) is far more complex; it shows the multi-photon resonances of photon blockade \cite{bishop_etal_2009,carmichael_2015} displayed in the bottom-right panel of Fig.~\ref{fig:fig1}. Moreover, with the ratio of time-dependent coefficients eliminated from Eqs.~(\ref{eqn:bloch1}) and (\ref{eqn:bloch2}), we find the constraint
\begin{equation}
\expect{\sigma_z(\infty)}[\expect{\sigma_z(\infty)}+1]=-2\left|\expect{\sigma_-(\infty)}\right|^2.
\label{eqn:constraint}
\end{equation}
The solution constructed from Eq.~(\ref{eqn:master_equation_lorentzian_spectrum}) violates Eq.~(\ref{eqn:constraint}) by as much as 100\%, with the right-hand side close to zero when the left, as a function of $\Delta/\lambda$, takes its maximum absolute value. The \emph{form} of the master equation reported in Ref.~\cite{shen_etal_2014} is therefore incorrect, not only the expressions for its time-dependent coefficients.

\begin{figure}[t]
\begin{center}
\includegraphics[width=3.4in]{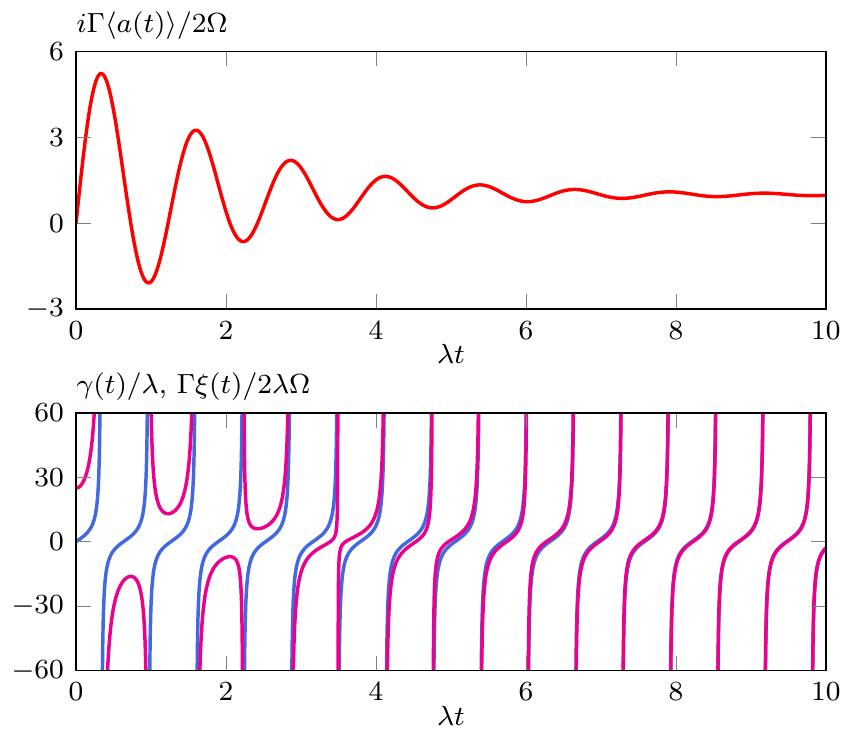}
\end{center}
\vskip-0.5cm
\caption{Peculiar asymptotic behavior of the time-dependent coefficients, $\gamma(t)$ and $\xi(t)$, in the time-local master equation of a driven boson system; for a Lorentzian spectral density with $(\Gamma/2\lambda)^{1/2}=5$ and $\Delta=\delta=0$. (Top) All expectations, e.g., $\langle a(t)\rangle$ (as shown), evolve to a steady state. (Bottom) Both $\gamma(t)$ (blue) and $\xi(t)$ (magenta) oscillate forever with periodic singularities.}
\label{fig:fig2}
\end{figure}

We conclude with an observation pertinent to the time-local description of driven systems in general. It appears, from Eqs.~(\ref{eqn:bloch1}) and (\ref{eqn:bloch2}), that driven systems may reach a far-from-equilibrium state with all operator averages steady but time-dependent
coefficients that still evolve; such packaging of the dynamics can appear decidedly odd from a physical point of view. Indeed, returning to a coupling strength $(\lambda\Gamma/2)^{1/2}$ much larger than the linewidth $\lambda$, and choosing $\Delta=\delta=0$, to take a simplest case, $\gamma(t)$ and $\xi(t)$ both oscillate forever in the long-time limit, as $\tan[(\lambda\Gamma/2)^{1/2}t]$---with periodic singularities. Of course Eqs.~(\ref{eqn:bloch1}) and (\ref{eqn:bloch2}) are incorrect for the steady state of the driven qubit; but the equivalent equation, $\gamma(t)\expect{a(\infty)}=-i\xi(t)$ [$\expect{\sigma_z(\infty)}\to-1$ in Eq.~(\ref{eqn:bloch1})], holds for the steady state of the driven boson system. This asymptotic behavior is therefore a feature of a formally correct---if decidedly odd---time-local ``repackaging'' of the dynamics. It speaks a word of caution for the physical interpretation of coefficients in the time-local formulation of non-Markovian dynamics. Figure~\ref{fig:fig2} illustrates the odd asymptotic behavior of a strongly coupled boson system driven to a far-from-equilibrium steady state.

We derived the time-local master equation for a driven boson system interacting with a boson environment by way of a time-local Heisenberg--Langevin equation. Our derivation is clear and direct: first, the time-nonlocal Heisenberg--Langevin equation is solved and the solution inverted; a time-local Heisenberg--Langevin equation can then be constructed by differentiating the solution and eliminating initial conditions; finally, taking expectations and using Wick's theorem to remove system-environment correlations, the time-local master equation is read off by inspection.

Our approach does not extend, however, to the driven qubit where the Heisenberg--Langevin equation is nonlinear and cannot be solved. Indeed, we showed that the tempting transcription, $a,a^\dagger\to\sigma_-,\sigma_+$ \cite{shen_etal_2014}, is generally incorrect, though valid for spontaneous emission and, as an approximation, if the excitation is weak. While it can account at some level for power broadening, it overlooks multiphoton resonances that derive from
the many-body excited states of the qubit coupled to its environment.

We also uncovered an oddity of the ``repackaged'' time-local dynamics in a far-from-equilibrium steady state, where although the density operator is steady, coefficients in the master equation oscillate forever.
\begin{acknowledgments}
The authors acknowledge helpful discussions with Jim Cresser, Arne Grimsmo, and Ricardo Guti\'errez-J\'auregui; also useful feedback from Francesco Petruccione and Ilya Sinayskiy. This work was supported by the Marsden Fund of the RSNZ.
\end{acknowledgments}


\end{document}